\documentclass[preprint, 12pt]{elsarticle}

\usepackage{amssymb}
\usepackage{graphicx}

\journal{Arxiv}

\begin{document}

\begin{frontmatter}

\title{Synchronised firing patterns in a random network of adaptive exponential
integrate-and-fire}

\author{F. S. Borges$^1$, P. R. Protachevicz$^1$, E. L. Lameu$^1$, 
R. C. Bonetti$^1$, K. C. Iarosz$^2$, I. L. Caldas$^2$, M. S. Baptista$^3$, 
A. M. Batista$^{1,2,4,*}$}
\address{$^1$P\'os-Gradua\c c\~ao em Ci\^encias/F\'isica, Universidade 
Estadual de Ponta Grossa, Ponta Grossa, PR, Brazil.}
\address{$^2$Instituto de F\'isica, Universidade de S\~ao Paulo, S\~ao Paulo, 
SP, Brazil.}
\address{$^3$Institute for Complex Systems and Mathematical Biology, 
Aberdeen, SUPA, UK.}
\address{$^4$Departamento de Matem\'atica e Estat\'istica, Universidade 
Estadual de Ponta Grossa, Ponta Grossa, PR, Brazil.}

\cortext[cor]{Corresponding author: antoniomarcosbatista@gmail.com}

\date{\today}

\begin{abstract}
We have studied neuronal synchronisation in a random network of adaptive 
exponential integrate-and-fire neurons. We study how spiking or bursting 
synchronous behaviour appears as a function of the coupling strength
and the probability of connections, by constructing parameter spaces that
identify these synchronous behaviours from measurements of the inter-spike
interval and the calculation of the order parameter. Moreover, we verify the
robustness of synchronisaton by applying an external perturbation to each 
neuron. The simulations show that bursting synchronisation is more robust than 
spike synchronisation.
\end{abstract}


\end{frontmatter}


\section{Introduction}
 
The concept of synchronistion is based on the adjustment of rhythms of
oscillating systems due to their interaction \cite{pikovsky01}.
Synchronisation phenomenon was recognised by Huygens in the 17th century, time
when he performed experiments to understand this phenomenon \cite{bennett02}.
To date, several kinds of synchronisation among coupled systems were reported,
such as complete \cite{li16}, phase \cite{pereira07,batista10}, lag
\cite{huang14}, and collective almost synchronisation \cite{baptista12}.

Neuronal synchronous rhythms have been observed in a wide range of researches 
about cognitive functions \cite{wang10,hutcheon00}. Electroencephalography and 
magnetoencephalography studies have been suggested that neuronal 
synchronization in the gamma frequency plays a functional role for memories in 
humans \cite{axmacher06,fell11}. Steinmetz et al. \cite{steinmetz00}
investigated the synchronous behaviour of pairs of neurons in the secondary 
somatosensory cortex of monkey. They found that attention modulates 
oscillatory neuronal synchronisation in the somatosensory cortex. Moreover,
in the literature it has been proposed that there is a relationship between
conscious perception and synchronisation of neuronal activity \cite{hipp11}.

We study spiking and bursting synchronisation betwe\-en neuron in a neuronal 
network model. A spike refers to the action potential generated by a neuron 
that rapidly rises and falls \cite{lange08}, while bursting refers to a 
sequence of spikes that are followed by a quiescent time \cite{wu12}. It was 
demonstrated that spiking synchronisation is relevant to olfactory bulb 
\cite{davison01} and is involved in motor cortical functions \cite{riehle97}.
The characteristics and mechanisms of bursting synchronisation were studied in 
cultured cortical neurons by means of planar electrode array \cite{maeda95}. 
Jefferys $\&$ Haas discovered synchronised bursting of CA1 hippocampal 
pyramidal cells \cite{jefferys82}.
 
There is a wide range of mathematical models used to describe neuronal activity,
such as the cellular automaton \cite{viana14}, the Rulkov map
\cite{rulkov01}, and differential equations \cite{hodgkin52,hindmarsh84}. 
One of the simplest mathematical models and that is widely used to depict 
neuronal behaviour is the integrate-and-fire \cite{lapicque07}, which is 
governed by a linear differential equation. A more realistic version of it is 
the adaptive exponential integrate-and-fire (aEIF) model which we consider in
this work as the local neuronal activity of neurons in the network. The aEIF is
a two-dimensional integrate-and-fire model introduced by Brette $\&$ Gerstner 
\cite{brette05}. This model has an exponential spike mechanism with an 
adaptation current. Touboul $\&$ Brette \cite{touboul08} studied the 
bifurcation diagram of the aEIF. They showed the existence of the Andronov-Hopf
bifurcation and saddle-node bifurcations. The aEIF model can generate multiple
firing patterns depending on the parameter and which fit experimental data from
cortical neurons under current stimulation \cite{naud08}.

In this work, we focus on the synchronisation phenomenon in a randomly
connected network. This kind of network, also called Erd\"os-R\'enyi network
\cite{erdos59}, has nodes where each pair is connected according to a
probability. The random neuronal network was utilised to study oscillations in
cortico-thalamic circuits \cite{gelenbe98} and dynamics of network with
synaptic depression \cite{senn96}. We built a random neuronal network with
unidirectional connections that represent chemical synapses.   

We show that there are clearly separated ranges of parameters that lead to
spiking or bursting synchronisation. In addition, we analyse the robustness to
external perturbation of the synchronisation. We verify that bursting 
synchronisation is more robustness than spiking synchronisation. However,
bursting synchronisation requires larger chemical synaptic strengths, and
larger voltage potential relaxation reset to appear than those required for
spiking synchronisation.

This paper is organised as follows: in Section II we present the adaptive 
exponential integrate-and-fire model. In Section III, we introduce the neuronal 
network with random features. In Section IV, we analyse the behaviour of 
spiking and bursting synchronisation. In the last Section, we draw our 
conclusions.


\section{Adaptive exponential integrate-and-fire}

As a local dynamics of the neuronal network, we consider the adaptive 
exponential integrate-and-fire (aEIF) model that consists of a system of two 
differential equations \cite{brette05} given by 
\begin{eqnarray}\label{eqIF}
C \frac{d V}{d t} & = & - g_L (V - E_L) + {\Delta}_T 
\exp \left(\frac{V - V_T}{{\Delta}_T} \right)   \nonumber \\
& & +I-w  , \nonumber \\
\tau_w \frac{d w}{d t} & = & a (V - E_L) - w,
\end{eqnarray}
where $V(t)$ is the membrane potential when a current $I(t)$ is injected, $C$ 
is the membrane capacitance, $g_L$ is the leak conductance, $E_L$ is the 
resting potential, $\Delta_T$ is the slope factor, $V_T$ is the threshold 
potential, $w$ is an adaptation variable, $\tau_w$ is the time constant, and 
$a$ is the level of subthreshold adaptation. If $V(t)$ reaches the threshold
$V_{\rm{peak}}$, a reset condition is applied: $V\rightarrow V_r$ and
$w\rightarrow w_r=w+b$. In our simulations, we consider $C=200.0$pF, 
$g_L=12.0$nS, $E_L=-70.0$mV, ${\Delta}_T=2.0$mV, $V_T=-50.0$mV, $I=509.7$pA, 
$\tau_w=300.0$ms, $a=2.0$nS, and $V_{\rm{peak}}=20.0$mV \cite{naud08}. 

\begin{figure}[hbt]
\centering
\includegraphics[height=7cm,width=10cm]{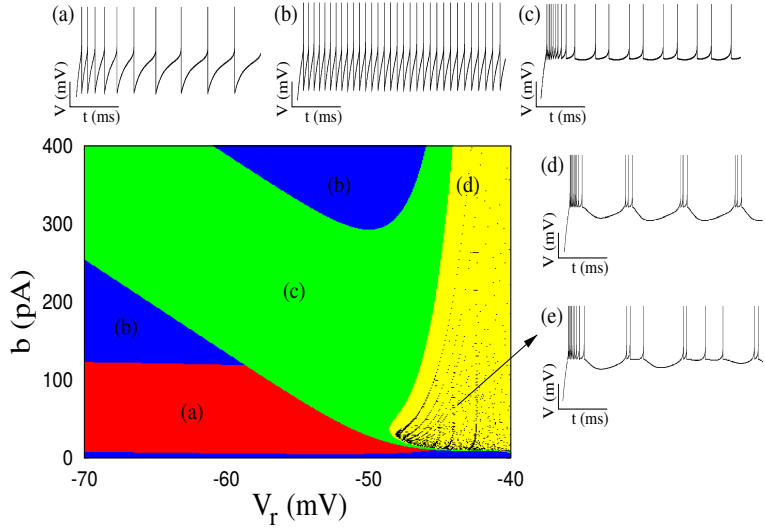}
\caption{(Colour online) Parameter space for the firing patterns as a function
of the reset parameters $V_r$ and $b$. (a) Adaptation in red, (b) tonic spiking
in blue, (c) initial bursting in green, (d) regular bursting in yellow, and (e) 
irregular in black.}
\label{fig1}
\end{figure}

The firing pattern depends on the reset parameters $V_r$ and $b$. Table
\ref{table1} exhibits some values that generate five different firing patterns
(Fig. \ref{fig1}). In Fig. \ref{fig1} we represent each firing pattern with a
different colour in the parameter space $b\times V_r$: adaptation in red, tonic
spiking in blue, initial bursting in green, regular bursting in yellow, and
irregular in black. In Figs. \ref{fig1}a, \ref{fig1}b, and \ref{fig1}c we
observe adaptation, tonic spiking, and initial burst pattern, respectively, due
to a step current stimulation. Adaptation pattern has increasing inter-spike
interval during a sustained stimulus, tonic spiking pattern is the simplest
regular discharge of the action potential, and the initial bursting pattern 
starts with a group of spikes presenting a frequency larger than the steady 
state frequency. The membrane potential evolution with regular bursting is 
showed in Fig. \ref{fig1}d, while Fig. \ref{fig1}e displays irregular pattern.

\begin{table}[htbp]
\caption{Reset parameters.}
\centering
\begin{tabular}{c c c c c}
\hline
Firing patterns & Fig. & b (pA) & $V_r$ (mV) & Layout \\ \hline
adaptation &\ref{fig1}(a) & 60.0 & -68.0 & red \\ 
tonic spiking & \ref{fig1}(b) & 5.0 & -65.0 &  blue\\ 
initial burst & \ref{fig1}(c) & 35.0 & -48.8 & green \\ 
regular bursting & \ref{fig1}(d) & 40.0 & -45.0 &  yellow\\ 
irregular & \ref{fig1}(e) & 41.2 & -47.4 & black \\ \hline
\end{tabular}
\label{table1}
\end{table}

\begin{figure}[hbt]
\centering
\includegraphics[height=7cm,width=9cm]{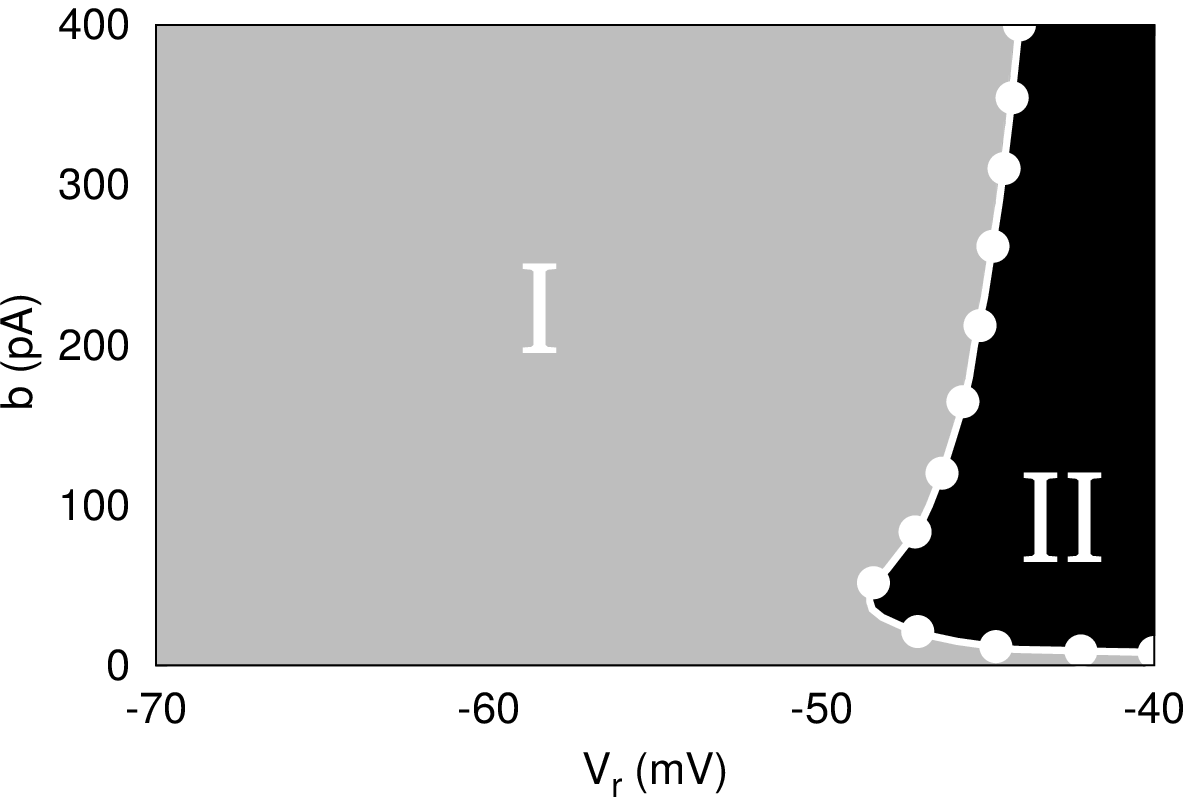}
\caption{Parameter space for the firing patterns as a function of the reset 
parameters $V_r$ and $b$. Spike pattern in region I ($\rm{CV}<0.5$) and 
bursting pattern in region II ($\rm{CV}\geq 0.5$) are separated by white 
circles.}
\label{fig2}
\end{figure}

As we have interest in spiking and bursting synchronisation, we separate the 
parameter space into a region with spike and another with bursting patterns 
(Fig. \ref{fig2}). To identify these two regions of interest, we use the 
coefficient of variation (CV) of the neuronal inter-spike interval (ISI), that 
is given by
\begin{eqnarray}\label{CV}
 {\rm CV}=\frac{{\sigma}_{\rm{ISI}}}{\rm{\overline{ISI}}},
\end{eqnarray}
where ${\sigma}_{\rm{ISI}}$ is the standard deviation of the ISI normalised by 
the mean $\bar{\rm ISI}$ \cite{gabbiani98}. Spiking patterns produce
$\rm{CV}<0.5$. Parameter regions that represent the neurons firing with spiking
pattern are denoted by gray colour in Fig. \ref{fig2}. Whereas, the black
region represents the bursting patterns, which results in $\rm{CV} \geq 0.5$.

\section{Spiking or bursting synchronisation}

In this work, we constructed a network where the neurons are randomly connected
\cite{erdos59}. Our network is given by
\begin{eqnarray}\label{eqIFrede}
C \frac{d V_i}{d t} & = & - g_L (V_i - E_L) + {\Delta}_T \; \rm{exp} 
\left(\frac{V_i - V_T}{{\Delta}_T} \right)  \nonumber \\
& + & I_i - w_i + g_{\rm{ex}} (V_{\rm{ex}} - V_i) \sum_{j=1}^N A_{ij} s_j + \Gamma_i, 
\nonumber \\
\tau_w \frac{d w_i}{d t} & = & a_i (V_i - E_L) - w_i, \nonumber \\
\tau_{\rm{ex}} \frac{d s_i}{d t} & = & - s_i.
\end{eqnarray}
where $V_i$ is the membrane potential of the neuron $i$, $g_{\rm{ex}}$ is the 
synaptic conductance, $V_{\rm{ex}}$ is the synaptic reversal potential, 
$\tau_{\rm{ex}}$ is the synaptic time constant, $s_i$ is the synaptic weight,
$A_{ij}$ is the adjacency matrix, $\Gamma_i$ is the external perturbation, and 
$a_i$ is randomly distributed in the interval $[1.9,2.1]$.

\begin{figure}[hbt]
\centering
\includegraphics[height=6cm,width=9cm]{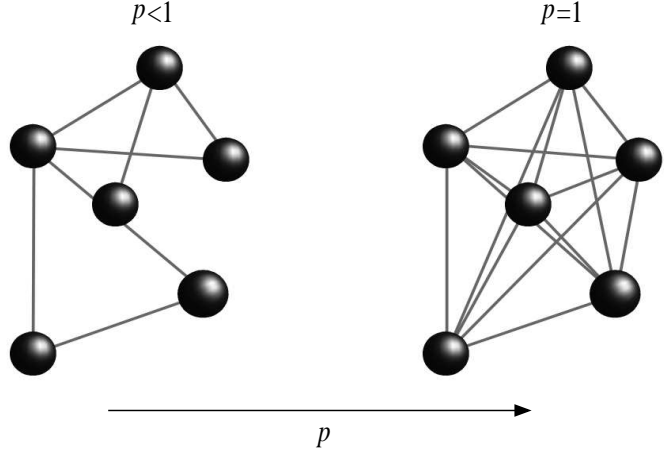}
\caption{Schematic representation of the neuronal network where the neurons
are connected according to a probability $p$.}
\label{fig3}
\end{figure}

The schematic representation of the neuronal network that we have considered
is illustrated in Fig \ref{fig3}. Each neuron is randomly linked to other
neurons with a probability $p$ by means of directed connections. When $p$ is 
equal to 1, the neuronal network becames an all-to-all network. A network with 
this topology was used by Borges et al. \cite{borges16} to study the effects 
of the spike timing-dependent plasticity on the synchronisation in a 
Hodgkin-Huxley neuronal network.

A useful diagnostic tool to determine synchronous behaviour is the complex
phase order parameter defined as \cite{kuramoto03}
\begin{equation}
z(t)=R(t)\exp({\rm i}\Phi(t))\equiv\frac{1}{N}\sum_{j=1}^{N}\exp({\rm i}\psi_{j}),
\end{equation}
where $R$ and $\Phi$ are the amplitude and angle of a centroid phase vector,
respectively, and the phase is given by
\begin{equation}
\psi_{j}(t)=2\pi m+2\pi\frac{t-t_{j,m}}{t_{j,m+1}-t_{j,m}},
\end{equation}
where $t_{j,m}$ corresponds to the time when a spike $m$ ($m=0,1,2,\dots$) of a 
neuron $j$ happens ($t_{j,m}< t < t_{j,m+1}$). We have considered the beginning 
of the spike when $V_j>-20$mV. The value of the order parameter magnitude goes 
to 1 in a totally synchronised state. To study the neuronal synchronisation of 
the network, we have calculated the time-average order-parameter, that is given
by 
\begin{equation}
\overline{R}=\frac{1}{t_{\rm fin}-{t_{\rm ini}}}\sum_{t_{\rm ini}}^{t_{\rm fin}}R(t),
\end{equation} 
where $t_{\rm fin}-t_{\rm ini}$ is the time window for calculating $\bar{R}$.

\begin{figure}[hbt]
\centering
\includegraphics[height=11cm,width=10cm]{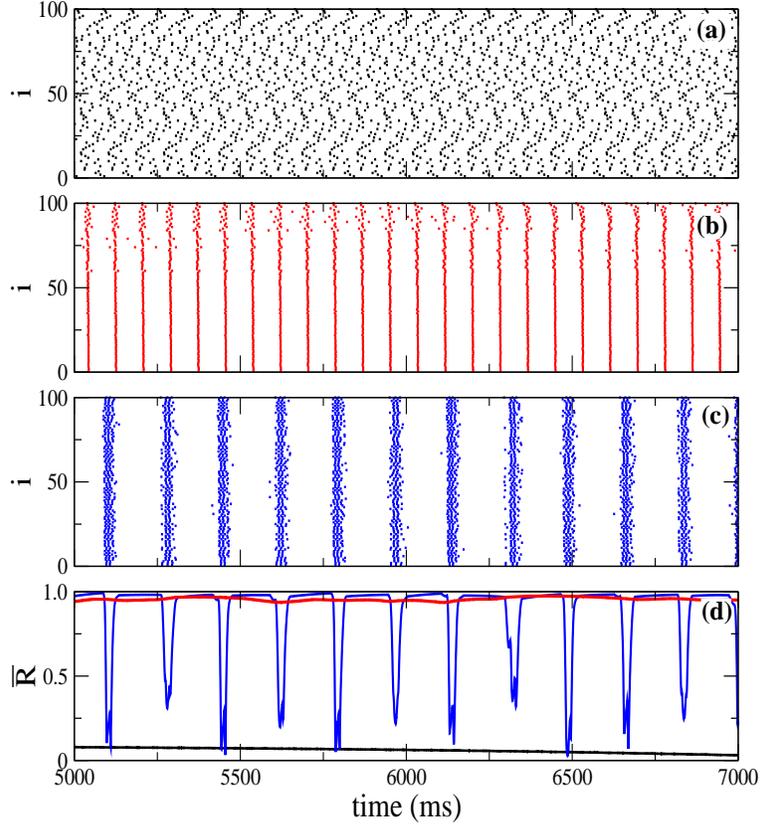}
\caption{(Colour online) Raster plot for (a) $g_{\rm ex}=0.02$nS, (b) 
$g_{\rm ex}=0.19$nS, and (c) $g_{\rm ex}=0.45$nS, considering $V_r = -58$mV, 
$p=0.5$,  and $b=70$pA. In (d) the order parameter is computed for 
$g_{\rm ex}=0.02$nS (black line), $g_{\rm ex}=0.19$nS (red line), and 
$g_{\rm ex}=0.19$nS (blue line).}
\label{fig4}
\end{figure}

Figs. \ref{fig4}a, \ref{fig4}b, and \ref{fig4}c show the raster plots for
$g_{\rm ex}=0.02$nS, $g_{\rm ex}=0.19$nS, and $g_{\rm ex}=0.45$nS, respectively, 
considering $V_r=-58$mV, $p=0.5$, and $b=70$pA, where the dots correspond to
the spiking activities generated by neurons. For $g_{\rm ex}=0.02$nS (Fig.
\ref{fig4}a) the network displays a desynchonised state, and as a result,
the order parameter values are very small (black line in Fig. \ref{fig4}d).
Increasing the synaptic conductance for $g_{\rm ex}=0.19$nS, the neuronal network
exhibits spike synchronisation (Fig. \ref{fig4}b) and the order parameter values
are near unity (red line in Fig. \ref{fig4}d). When the network presents 
bursting synchronisation (Fig. \ref{fig4}c), the order parameter values vary
between $R\approx 1$ and $R\ll 1$ (blue line in Fig. \ref{fig4}d). $R\ll 1$ to
the time when the neuron are firing.

\begin{figure}[hbt]
\centering
\includegraphics[height=7cm,width=9cm]{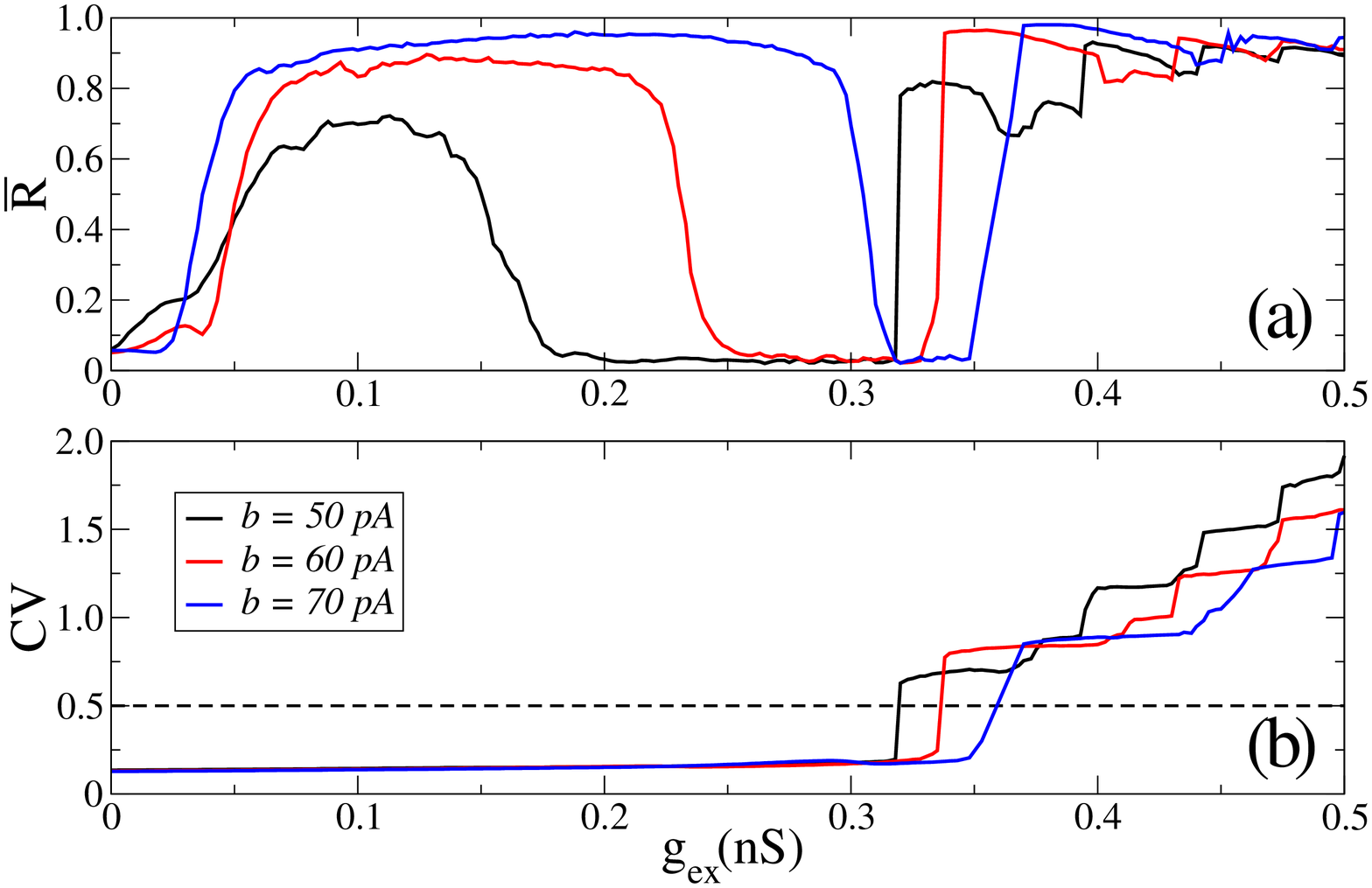}
\caption{(Colour online) (a) Time-average order parameter and (b) CV for 
$V_r=-58$mV, $p=0.5$, $b=50$pA (black line), $b=60$pA (red line), and $b=70$pA
(blue line).}
\label{fig5}
\end{figure}

In Fig. \ref{fig5}a we show ${\bar R}$ as a function of $g_{\rm ex}$ for 
$p=0.5$, $b=50$pA (black line), $b=60$pA (red line), and $b=70$pA (blue line). 
The three results exhibit strong synchronous behaviour (${\bar R}>0.9$) for 
many values of $g_{\rm ex}$ when $g_{\rm ex}\gtrsim 0.4$nS . However, for 
$g_{\rm ex}\lesssim 0.4$nS, it is possible to see synchronous behaviour only for
$b=70$pA in the range $0.15{\rm nS}<g_{\rm ex}<0.25{\rm nS}$. In addition, we 
calculate the coefficient of variation (CV) to determine the range in 
$g_{\rm ex}$ where the neurons of the network have spiking or bursting behaviour 
(Fig. \ref{fig5}b). We consider that for CV$<0.5$ (black dashed line) the 
neurons exhibit spiking behaviour, while for CV$\geq 0.5$ the neurons present 
bursting behaviour. We observe that in the range 
$0.15{\rm nS}<g_{\rm ex}<0.25{\rm nS}$ for $b=70$pA there is spiking 
sychronisation, and bursting synchronisation for $g_{\rm ex}\gtrsim 0.4$nS. 

\section{Parameter space of synchronisation}

\begin{figure}[hbt]
\centering
\includegraphics[height=6cm,width=9cm]{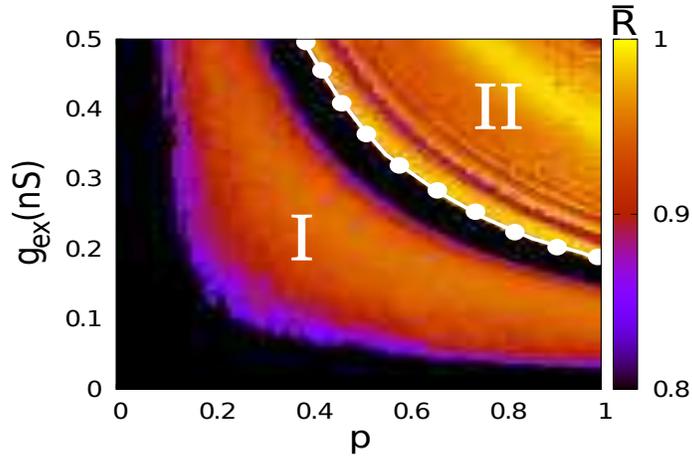}
\caption{(Colour online) $g_{\rm ex} \times p$ for $V_r=-58$mV and $b=70$pA,
where the colour bar represents the time-average order parameter. The regions I
(spike patterns) and II (bursting patterns) are separated by the white line
with circles.}
\label{fig6}
\end{figure}

The synchronous behaviour depends on the synaptic conductance and the 
probability of connections. Fig. \ref{fig6} exhibits the time-averaged order 
parameter in colour scale as a function of $g_{\rm ex}$ and $p$. We verify a
large parameter region where spiking and bursting synchronisation is strong,
characterised by ${\bar R}>0.9$. The regions I and II correspond to spiking and
bursting patterns, respectively, and these regions are separated by a white 
line with circles. We obtain the regions by means of the coefficient of 
variation (CV). There is a transition between region I and region II, where 
neurons initially synchronous in the spike, loose spiking synchronicity to give 
place to a neuronal network with a regime of bursting synchronisation. 

\begin{figure}[hbt]
\centering
\includegraphics[height=12cm,width=7cm]{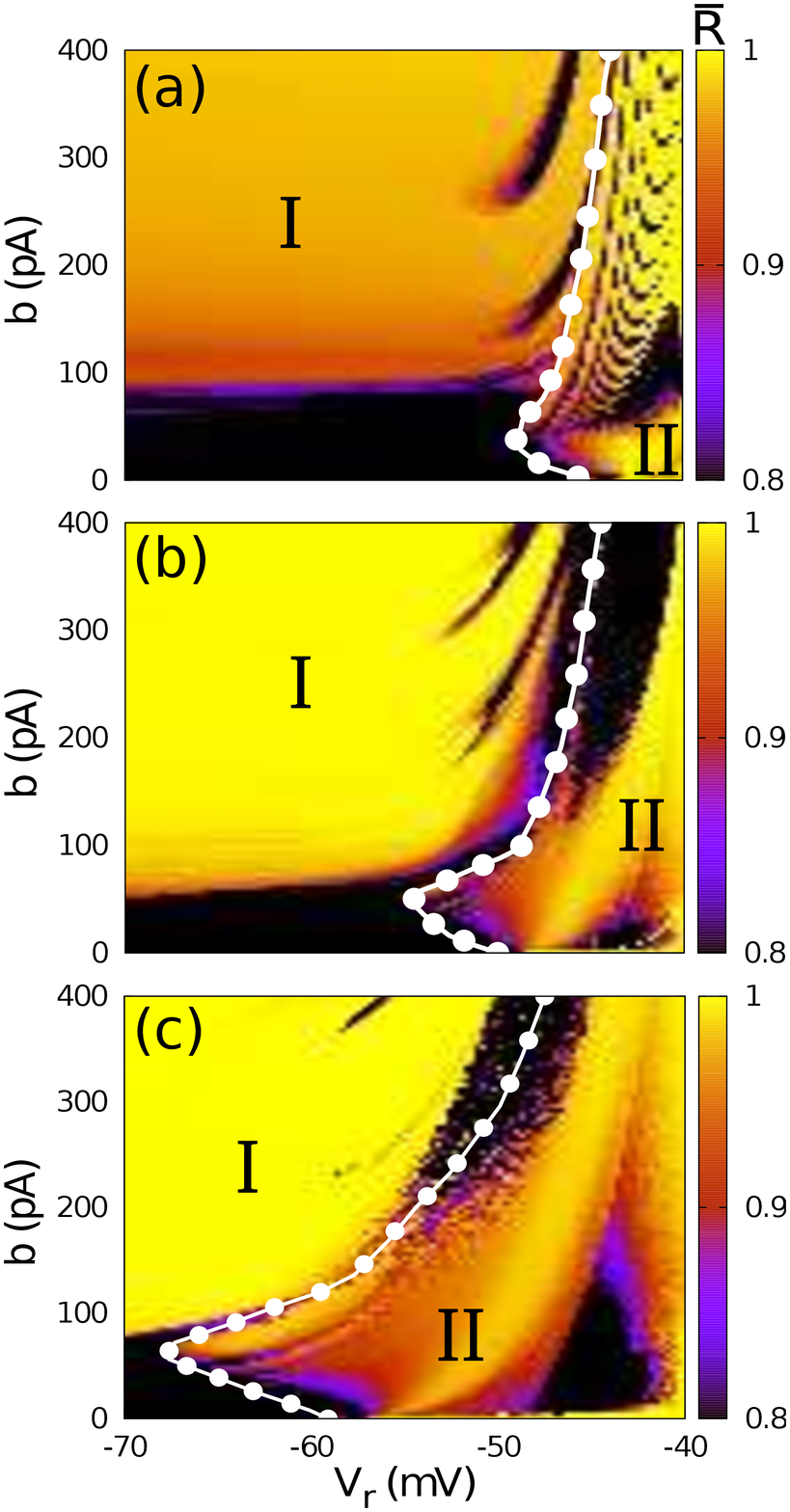}
\caption{(Colour online) Parameter space $b \times V_r$ for $p=0.5$, $\gamma=0$
(a) $g_{\rm ex}=0.05$nS, (b) $g_{\rm ex}=0.25$nS, and (c) $g_{\rm ex}=0.5$nS, where 
the colour bar represents the time-average order parameter. The regions I
(spike patterns) and II (bursting patterns) are separated by white circles.}
\label{fig7}
\end{figure}

We investigate the dependence of spiking and bursting synchronisation on the 
control parameters $b$ and $V_r$. To do that, we use the time average order 
parameter and the coefficient of variation. Figure \ref{fig7} shows that the
spike patterns region (region I) decreases when $g_{\rm ex}$ increases. This 
way, the region I for $b<100$pA and $V_r=-49$mV of parameters leading to no
synchronous behaviour (Fig. \ref{fig7}a), becomes a region of parameters that
promote synchronised bursting (Fig. \ref{fig7}b and \ref{fig7}c). However, a
large region of desynchronised bursting appears for $g_{\rm ex}=0.25$nS about
$V_r=-45$mV and $b>100$pA in the region II (Fig. \ref{fig7}b). For
$g_{\rm ex}=0.5$nS, we see, in Fig. \ref{fig7}c, three regions of desynchronous
behaviour, one in the region I for $b<100$pA, other in region II for $b<200$pA,
and another one is located around the border (white line with circles) between
regions I and II for $b>200$pA.

It has been found that external perturbations on neuronal networks not only can
induce synchronous behaviour \cite{baptista06,zhang15}, but also can suppress 
synchronisation \cite{lameu16}. Aiming to study the robustness to perturbations
of the synchronous behaviour, we consider an external perturbation $\Gamma_i$
(\ref{eqIFrede}). It is applied on each neuron $i$ with an average time 
interval of about $10$ms and with a constant intensity $\gamma$ during $1$ms.

\begin{figure}[hbt]
\centering
\includegraphics[height=12cm,width=7cm]{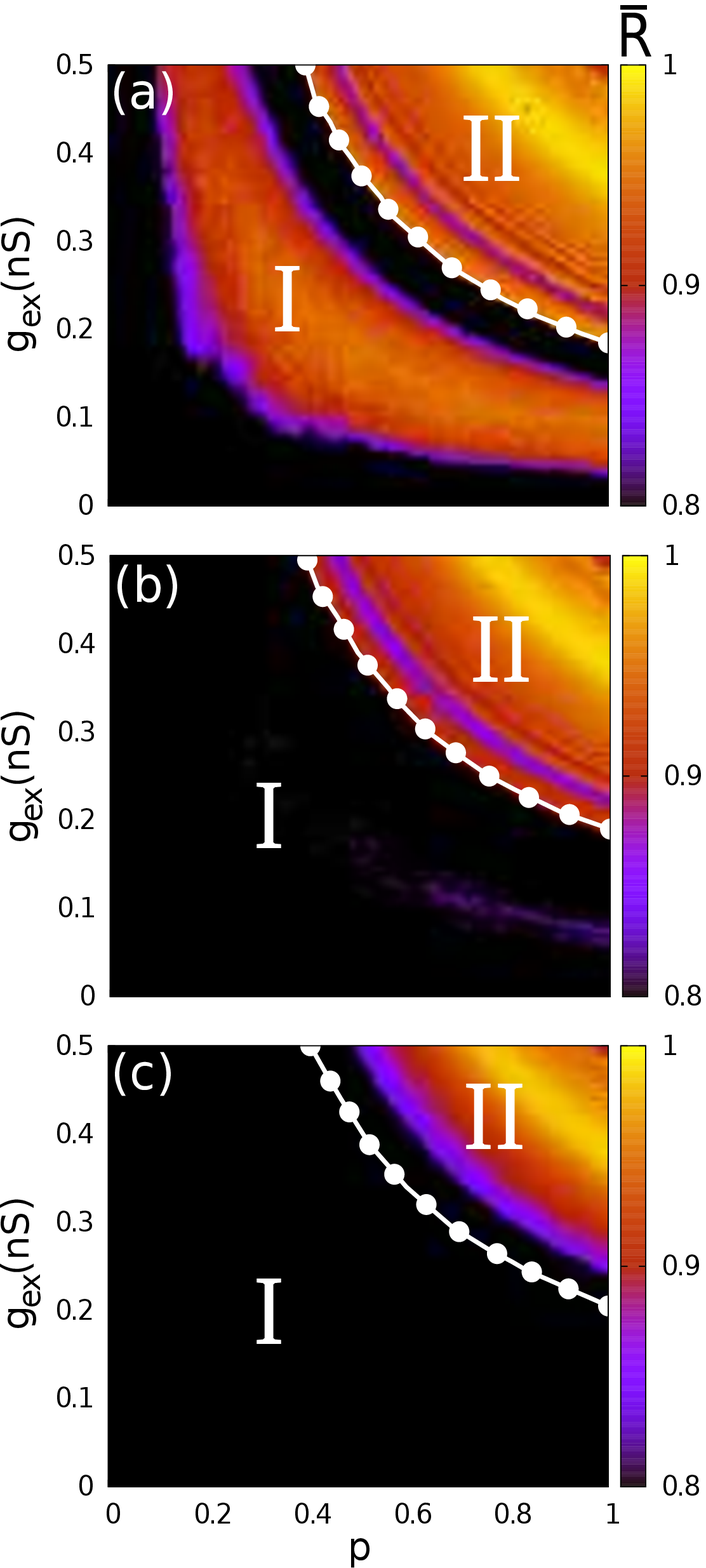}
\caption{(Colour online) $g_{\rm ex} \times p$ for $V_r=-58$mV, $b=70$pA,
(a) $\gamma=250$pA, (b) $\gamma=500$pA, and (c) $\gamma=1000$pA.}
\label{fig8}
\end{figure}

Figure \ref{fig8} shows the plots $g_{\rm ex} \times p$ for $\gamma>0$, where 
the regions I and II correspond to spiking and bursting patterns, respectively, 
separated by white line with circles, and the colour bar indicates the
time-average order parameter values. In this Figure, we consider $V_r=-58$mV,
$b=70$pA, (a) $\gamma=250$pA, (b) $\gamma=500$pA, and (c) $\gamma=1000$pA. For
$\gamma=250$pA (Fig. \ref{fig8}a) the perturbation does not suppress spike
synchronisation, whereas for $\gamma=500$pA the synchronisation is
completely suppressed in region I (Fig. \ref{fig8}b). In Fig. \ref{fig8}c, we
see that increasing further the constant intensity for $\gamma=1000$pA, the
external perturbation suppresses also bursting synchronisation in region II.
Therefore,the synchronous behavior in region II is more robustness to
perturbations than in the region I, due to the fact that the region II is in a
range with high $g_{\rm ex}$ and $p$ values, namely strong coupling and high
connectivity.

\begin{figure}[hbt]
\centering
\includegraphics[height=5cm,width=7cm]{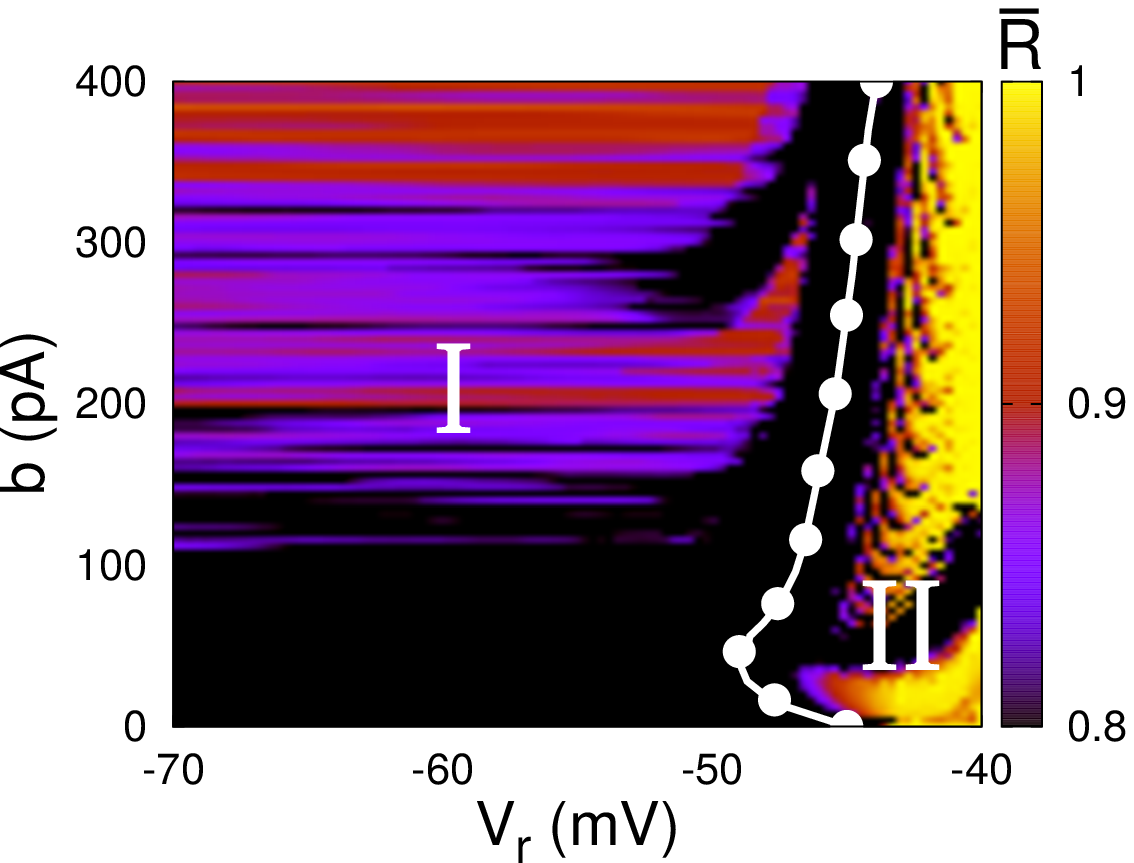}
\caption{(Colour online) $b \times V_r$ for $g_{\rm ex}=0.05$nS, $p=0.5$, and
$\gamma=500$pA, where the colour bar represents the time-average order 
parameter. The regions I (spike patterns) and II (bursting patterns) are 
separated by white line with circles.}
\label{fig9}
\end{figure}

In order to understand the perturbation effect on the spike and bursting
patterns, we consider the same values of $g_{\rm ex}$ and $p$ as Fig.
\ref{fig7}a. Figure \ref{fig9} exhibits the space parameter $b\times V_r$,
where $\gamma$ is equal to $500$pA. The external perturbation suppresses  
synchronisation in the region I, whereas we observe synchronisation in 
region II. The synchronous behaviour in region II can be suppressed if the
constant intensity $\gamma$ is increased. Therefore, bursting synchronisation
is more robustness to perturbations than spike synchronisation.

\section{Conclusion}

In this paper, we studied the spiking and bursting synchronous behaviour in a
random neuronal network where the local dynamics of the neurons is given by the
adaptive exponential integrate-and-fire (aEIF) model. The aEIF model can exhibit
different firing patterns, such as adaptation, tonic spiking, initial burst,
regular bursting, and irregular bursting. 

In our network, the neurons are randomly connected according to a probability.
The larger the probability of connection, and the strength of the synaptic
connection, the more likely is to find bursting synchronisation.

It is possible to suppress synchronous behaviour by means of an external
perturbation. However, synchronous behaviour with higher values of 
$g_{\rm ex}$ and $p$, which typically promotes bursting synchronisation, are more
robust to perturbations, then spike synchronous behaviour appearing for
smaller values of these parameters. We concluded that bursting synchronisation 
provides a good environment to transmit information when neurons are stron\-gly 
perturbed (large $\Gamma$).


\section*{Acknowledgements}
This study was possible by partial financial support from the following 
Brazilian government agencies: CNPq, CAPES, and FAPESP (2011/19296-1 and
2015/07311-7). We also wish thank Newton Fund and COFAP.


\end{document}